\documentclass{article}
\usepackage{spconf, amsmath, amssymb, mathtools, graphicx, tikz, pgfplots, xcolor, stmaryrd}
\usepackage{nohyperref}
\usepackage[belowskip=-0.4cm,aboveskip=0.2cm]{caption}
\usetikzlibrary{arrows,calc,plotmarks,patterns,positioning,shapes}
\usepackage[numbers,sort&compress]{natbib}

\AtBeginDocument{
  \addtolength\abovedisplayskip{-2.2pt}
  \addtolength\belowdisplayskip{-2.2pt}
}

\title{Super-resolution from Short-Time Fourier Transform Measurements}

\name{
	C\'eline Aubel, David Stotz, and Helmut B\"olcskei \thanks{The authors would like to thank J.-P.~Kahane for answering questions on results in~\cite{Kahane2011}, M.~Lerjen for his help with simulations, and C.~Chenot for inspiring discussions.}
}
\address{Dept.~IT\&EE, ETH Zurich, Switzerland \\ \{aubelc, dstotz, boelcskei\}@nari.ee.ethz.ch}

\newtheorem{thm}{Theorem}

\DeclareMathOperator{\sinc}{sinc}
\DeclareMathOperator{\real}{Re}

\renewcommand{\Re}[1]{\real\left\{#1\right\}}

\DeclareMathOperator{\supp}{supp}

\newcommand{\abs}[1]{\left|#1\right|}
\DeclareMathOperator{\tr}{Tr}

\renewcommand{\triangleq}{\vcentcolon=}

\DeclareMathOperator*{\minimize}{minimize\ }
\DeclareMathOperator*{\maximize}{maximize\ }
\DeclareMathOperator{\st}{\ subject\ to\ }
\DeclareMathOperator{\gaborProb}{(SR)}
\DeclareMathOperator{\dualGaborProb}{(PD-SR)}

\newcommand{\C}{\mathbb{C}}
\newcommand{\R}{\mathbb{R}}
\newcommand{\Z}{\mathbb{Z}}

\newcommand{\torus}{\mathbb{T}}

\renewcommand{\leq}{\leqslant}
\renewcommand{\geq}{\geqslant}

\newcommand{\meas}{\mu}
\newcommand{\measProb}{\nu}

\newcommand{\measSpace}[1]{\mathcal{M}(#1)}

\newcommand{\group}{G}

\newcommand{\freq}{f}
\newcommand{\borelAlgebra}[1]{\mathcal{B}(#1)}
\newcommand{\measSupport}{T}
\newcommand{\measSupportIdx}{\Omega}
\newcommand{\measAmplitude}[1]{a_{#1}}
\newcommand{\measSpikeIdx}{\ell}

\newcommand{\dmeas}[2]{#1\langle#2\rangle}

\newcommand{\innerProd}[2]{\left\langle #1, #2 \right\rangle}
\DeclareMathOperator*{\tv}{TV}
\newcommand{\normTV}[1]{\left\|#1\right\|_{\tv}}

\newcommand{\normInfty}[1]{\left\|#1\right\|_{\infty}}

\newcommand{\measurements}{y}
\newcommand{\gaborOp}{\mathcal{A}_\window}
\newcommand{\window}{g}
\newcommand{\autocorrelation}{G}
\newcommand{\STFT}{\mathcal{V}_\window}

\begin{document}

\maketitle

\begin{abstract}
	While spike trains are obviously not band-limited, the theory of super-resolution tells us that perfect recovery of unknown spike locations and weights from low-pass Fourier transform measurements is possible provided that the minimum spacing, $\Delta$, between spikes is not too small. Specifically, for a cutoff frequency of $\freq_c$, Donoho~\cite{Donoho1991} shows that exact recovery is possible if $\Delta > 1/\freq_c$, but does not specify a corresponding recovery method.
On the other hand, Cand\`es and Fernandez-Granda~\cite{Candes2012} provide a recovery method based on convex optimization, which provably succeeds as long as $\Delta > 2/\freq_c$. 
In practical applications one often has access to windowed Fourier transform measurements, i.e., short-time Fourier transform (STFT) measurements, only.
In this paper, we develop a theory of super-resolution from STFT measurements, and we propose a method that provably succeeds in recovering spike trains from STFT measurements provided that $\Delta > 1/\freq_c$. 
\end{abstract}

\begin{keywords}
\hspace{-0.17cm}Super-resolution, inverse problems in measure spaces, short-time Fourier transform.
\end{keywords}

\section{Introduction}
\label{section: introduction}

The recovery of spike trains with unknown spike locations and weights from low-pass Fourier measurements, commonly referred to as super-resolution, has been a topic of long-standing interest~\cite{Logan1965, Logan1977, Beurling1966, Beurling1989-1, Beurling1989-2, Donoho1992, Vetterli2002, Dragotti2007}, with recent focus on $\ell^1$-minimization-based recovery techniques~\cite{Levy1981, Chen2001, Candes2012}.
It was recognized in~\cite{Donoho1991, Bredies2012, Castro2012, Candes2012, Duval2013} that a measure-theoretic formulation of the super-resolution problem in continuous-time not only leads to a clean mathematical framework, but also to results that are ``grid-free''~\cite{Tang2012}. This theory is inspired by Beurling's seminal work on the ``balayage'' of measures in Fourier transforms~\cite{Beurling1966, Beurling1989-1}, and on interpolation using entire functions of exponential type~\cite{Beurling1989-2}. Specifically, Castro and Gamboa~\cite{Castro2012}, and Cand\`es and Fernandez-Granda~\cite{Candes2012}, propose to solve a total variation minimization problem for recovering a discrete measure (modeling the spike train) from low-pass Fourier measurements. Despite its infinite-dimensional nature, this optimization problem can be solved exploiting Fenchel duality, as described in~\cite{Bredies2012, Candes2012, Duval2013}. Concretely, it is shown in~\cite{Castro2012, Candes2012} that the analysis of the Fenchel dual problem leads to an interpolation problem, which can be solved explicitly as long as the elements in the support of the measure to be recovered are separated by at least $2/\freq_c$, where $\freq_c$ represents the cutoff frequency of the low-pass measurements. Donoho~\cite{Donoho1991} proves that a separation of $1/\freq_c$ is sufficient for perfect recovery, but does not provide a concrete method for reconstructing the measure. Kahane~\cite{Kahane2011} shows that recovery is possible if the minimum separation between spikes is at least $\frac{5}{\freq_c}\sqrt{\log(1/\freq_c)}$, but this result suffers from a log-factor penalty. 

\textit{Contributions:} In practical applications one often has access to windowed Fourier transform, i.e., short-time Fourier transform (STFT), measurements only. It is therefore of interest to develop a theory of super-resolution from STFT measurements. This is precisely the goal of the present paper. Inspired by~\cite{Donoho1991, Bredies2012, Castro2012, Candes2012, Duval2013, AuYeung2013}, we use a measure-theoretic formulation and consider the continuous-time case. Our main result shows that exact recovery through convex optimization is possible, for a Gaussian window function, provided that the minimum separation between points in the measure support exceeds $1/\freq_c$.

\textit{Notation and preparatory material:} The complex conjugate of $z \in \C$ is denoted by $\overline{z}$. The derivative of the function $\varphi$ is designated by $\varphi'$. The $\sinc$ function is defined as $\sinc(t) \triangleq \sin(t)/t$ for all $t \neq 0$ and $\sinc(0) = 1$. Uppercase boldface letters stand for matrices. The entry in the $k$th row and $\ell$th column of the matrix $\mathbf{M}$ is $m_{k,\ell}$. The superscript $^H$ stands for Hermitian transposition. For matrices $\mathbf{X}, \mathbf{Y} \in \C^{M \times N}$, we write $\innerProd{\mathbf{X}}{\mathbf{Y}} \triangleq \Re{\tr(\mathbf{Y}^H\mathbf{X})}$ for their real inner product.
General linear operators are designated by uppercase calligraphic letters. If $X$ and $Y$ are topological vector spaces, and $X^*$ and $Y^*$ their topological duals, the adjoint of the linear operator $\mathcal{L} \colon X \rightarrow Y$ is denoted by $\mathcal{L}^* \colon Y^* \rightarrow X^*$.
The set of all solutions of an optimization problem $(\mathrm{P})$ is denoted by $\mathrm{Sol}\{(\mathrm{P})\}$.
For a measure space $(X, \Sigma, \meas)$ and a measurable function $\varphi \colon X \rightarrow \C$, we write $\int_X \varphi(x)\dmeas{\meas}{x}$ for the integration of $\varphi$ with respect to $\meas$, where we set $\mathrm{d}x \triangleq \dmeas{\lambda}{x}$ if $\lambda$ is the Lebesgue measure. 
For $p \in [1, \infty)$, $L^p(X, \Sigma, \meas)$ denotes the space of all functions $\varphi \colon X \rightarrow \C$ such that $\|\varphi\|_{L^p} \triangleq \left(\int_X \abs{\varphi(x)}^p\dmeas{\meas}{x}\right)^{1/p} < \infty$. The space $L^\infty(X, \Sigma, \meas)$ contains all functions $\varphi \colon X \rightarrow \C$ such that $\|\varphi\|_{L^\infty} \triangleq \inf\{C > 0 \colon \abs{\varphi(x)} \leq C \text{ for }\meas\text{-almost all }x \in X\} < \infty$. 
For functions $\varphi \in L^p(X, \Sigma, \meas)$ and $\psi \in L^q(X, \Sigma, \meas)$ with $p, q \in [1, \infty]$ satisfying $1/p+1/q = 1$, we set $\innerProd{\varphi}{\psi} \triangleq \Re{\int_X \varphi(x)\overline{\psi(x)}\dmeas{\meas}{x}}$.
For a separable locally compact metric abelian group $\group$ (e.g., the additive group $\R$ or the torus $\torus \triangleq \R/\Z$ endowed with the natural topology), we write $L^p(G)$ in the particular case where $\Sigma$ is the Borel $\sigma$-algebra of $G$ and $\meas$ the Haar measure on $\group$. We denote by $\borelAlgebra{\group}$ the Borel $\sigma$-algebra of $\group$ and by $\measSpace{\group}$ the space of all complex Radon measures on $(\group, \borelAlgebra{\group})$. For $t \in \group$, $\delta_t \in \measSpace{\group}$ designates the Dirac measure at $t$, which for $B \in \borelAlgebra{\group}$ is given by $\delta_t(B) = 1$ if $t \in B$ and $\delta_t(B) = 0$ otherwise. The support $\supp(\meas)$ of a complex Radon measure $\meas \in \measSpace{\group}$ is the largest closed set $C \subseteq \group$ such that for every open set $B \in \borelAlgebra{\group}$ satisfying $B \cap C \neq \emptyset$, it holds that $\meas(B \cap C) \neq 0$.
For every $\meas \in \measSpace{\group}$, the total variation (TV) of $\meas$ is defined as the measure $|\meas|$ satisfying
\begin{equation*}
	\forall B \in \borelAlgebra{\group}, \quad |\meas|(B) \triangleq \sup_{\pi \in \Pi(B)} \sum_{A \in \pi} |\meas(A)|,
\end{equation*}
where $\Pi(B)$ denotes the set of all partitions of $B$. The space $\measSpace{\group}$ can be equipped with the TV norm $\normTV{\meas} \triangleq |\meas|(\group)$.
Based on the Riesz representation theorem~{\cite[Thm.~6.19]{Rudin1987}}, $\measSpace{\group}$ can be characterized as the dual space of $C_0(\group)$. With $C_c(\group)$ the space of all complex-valued continuous functions on $\group$ whose support is compact, $C_0(\group)$ is the completion of $C_c(\group)$ relative to the metric defined by the supremum norm $\normInfty{\varphi} = \sup_{t \in \group} |\varphi(t)|$. By analogy with the real inner product in $L^2(\group)$, we define the real dual pairing of the measure $\meas \in \measSpace{\group}$ and the function $\varphi \in C_0(\group)$ as
$\innerProd{\meas}{\varphi} \triangleq \Re{\int_\group \overline{\varphi(t)}\dmeas{\meas}{t}}$.
We endow $\measSpace{\group}$ with the weak-* topology~\cite{Brezis1983}, i.e., the coarsest topology on $\measSpace{\group}$ for which every linear functional $\mathcal{L}_\varphi \colon \measSpace{\group} \rightarrow \R$ defined by $\meas \mapsto \mathcal{L}_\varphi(\meas) = \innerProd{\meas}{\varphi}$, with $\varphi \in C_0(\group)$, is continuous.

\section{Statement of the problem}
\label{section: statement of the problem}

We consider a complex Radon measure on $\R$ of the form
\begin{equation}
	\meas = \sum_{\measSpikeIdx \in \measSupportIdx} \measAmplitude{\measSpikeIdx} \delta_{t_\measSpikeIdx}.
	\label{eq: original measure}
\end{equation}	
The measure models a spike train, and is supported on the closed discrete set $\measSupport \triangleq \{t_\measSpikeIdx\}_{\measSpikeIdx \in \measSupportIdx} \subseteq \R$ with complex mass $\measAmplitude{\measSpikeIdx} \neq 0$ attached to the point $t_\measSpikeIdx$. The locations $t_\measSpikeIdx$ and the associated $\measAmplitude{\measSpikeIdx}$ are assumed unknown throughout the paper. We do, however, assume that the measure $\meas$ is known to be discrete. The set $\measSupportIdx$ is assumed to be countable, and we require that $\sum_{\measSpikeIdx \in \measSupportIdx} \abs{\measAmplitude{\measSpikeIdx}} < \infty$ for $\meas$ to be in $\measSpace{\R}$. Throughout the paper, $\meas$ designates exclusively the measure defined in~\eqref{eq: original measure}.

Suppose we obtain measurements of $\meas$ in the time-frequency domain in the form:
\begin{equation*}
	\measurements(\tau, \freq) = (\STFT\meas)(\tau, \freq)
\end{equation*}
for $\tau \in \R$ and $\freq \in B_{\freq_c} \triangleq \{\freq \in \R \colon \abs{\freq} \leq \freq_c\}$, where $\freq_c$ is the cutoff frequency and
\begin{equation}
	(\STFT\meas)(\tau, \freq) \triangleq \int_{\R} \overline{\window(t - \tau)}e^{-2\pi i\freq t}\dmeas{\meas}{t}
	\label{eq: STFT measure}
\end{equation}
denotes the STFT~\cite{Groechenig2000} of $\meas$ with respect to the window function $\window$.
Although our theory applies to more general window functions taken from the Schwartz space of rapidly decaying functions, for concreteness, we choose $\window$ to be Gaussian, i.e.,
\begin{equation*}
	\forall t \in \R, \quad \window(t) = \frac{1}{\sqrt{\sigma}}\exp\left(-\frac{t^2}{2\sigma^2}\right),
\end{equation*}
where $\sigma > 0$ is a parameter controlling the width of the window.
We wish to recover the measure~$\meas$ from the measurements $\measurements$ by solving the following optimization problem:
\begin{equation*}
	\gaborProb \quad \minimize_{\measProb \in \measSpace{\R}} \normTV{\measProb} \st \measurements = \gaborOp\measProb,
\end{equation*}
 where $\gaborOp \colon \measSpace{\R} \rightarrow L^1(\R^2)$ maps $\measProb \in \measSpace{\R}$ to the function $\varphi \in L^1(\R^2)$ such that
\begin{equation*}
	\forall (\tau, \freq) \in \R^2, \quad \varphi(\tau, \freq) = \begin{cases} (\STFT\measProb)(\tau, f), & f \in B_{\freq_c} \\
		0, & \text{otherwise}. \end{cases}
\end{equation*}
The motivation for considering time-frequency measurements is twofold. First, signals are often partitioned into short time segments and windowed for acquisition. Second, the frequency characteristics of the signal modeled by the measure $\meas$ often vary over time, i.e., the $t_\measSpikeIdx$, $\measSpikeIdx \in \measSupportIdx$, can be more packed in certain intervals, so that time-localized spectral information about $\meas$ will lead to improved reconstruction quality for the same frequency limitation.

\section{Reconstruction from complete measurements}
\label{section: reconstruction from complete measurements}

Before embarking on the problem of reconstructing $\meas$ by solving $\gaborProb$ for a given $\freq_c < \infty$, we need to convince ourselves that reconstruction is possible from complete measurements, i.e., for $\freq_c = \infty$. This should be evident as the STFT, defined for functions in $L^2(\R)$, is invertible~\cite{Groechenig2000}. The measure-theoretic STFT considered here is, however, non-standard.
While one can show that the STFT of a measure still determines the underlying measure uniquely in the sense that, for $\nu \in \measSpace{\R}$, $\STFT\nu = 0$ implies $\nu = 0$, to the best of our knowledge, no general inversion formula is available for complex Radon measures. Since $\meas$ is discrete, all we need to find is its support $\measSupport = \{t_\measSpikeIdx\}_{\measSpikeIdx \in \measSupportIdx}$ and the corresponding complex masses $\{\measAmplitude{\measSpikeIdx}\}_{\measSpikeIdx \in \measSupportIdx}$. Specifically,  it can be shown that $\meas$ can be recovered from its STFT $\STFT\meas$ according to
\begin{align*}
	\lim_{F \rightarrow \infty} \frac{1}{2F} &\int_{-F}^F\int_\R (\STFT\meas)(\tau, \freq)\window(t - \tau)e^{2\pi i\freq t}\mathrm{d}\tau\mathrm{d}\freq \notag\\
		&= \begin{cases} \measAmplitude{\measSpikeIdx}, & \text{if }\ t = t_\measSpikeIdx \\ 0, & \text{otherwise}. \end{cases}
\end{align*}

\section{Reconstruction from partial measurements}
\label{section: reconstruction from partial measurements}

Now, we consider the reconstruction of $\meas$ from band-limited STFT measurements using $\gaborProb$. Since the space $\measSpace{\R}$ is infinite-dimensional, the existence of a solution of $\gaborProb$ is delicate. It turns out, however, that relying  on the convexity of the TV norm $\normTV{\cdot}$ and on compactness of the unit ball $\{\measProb \in \measSpace{\group} \colon \normTV{\measProb} \leq 1\}$ with respect to the weak-* topology, the result provided in \cite[Cor.~3.20]{Brezis1983} ensures the existence of a solution of $\gaborProb$. Next, with the help of Fenchel duality theory~\cite[Chap.~4]{Borwein2005}, we derive necessary and sufficient conditions for $\meas$ to be the unique solution of $\gaborProb$.

\begin{thm}[Fenchel predual]
	\label{thm: fenchel dual gabor prob}
	The Fenchel predual problem of $\gaborProb$ is
	\begin{equation*}
		\dualGaborProb\qquad  \maximize_{c \in L^\infty(\R^2)} \innerProd{c}{\measurements} \st \normInfty{\gaborOp^*c} \leq 1.
	\end{equation*}
	In addition, the following equality holds 
	\begin{align}
		&\min\Big\{\normTV{\measProb} \colon \gaborOp\measProb = \measurements, \measProb \in \measSpace{\R}\Big\} \notag \\
		&\quad = \sup\Big\{\innerProd{c}{\measurements} \colon \normInfty{\gaborOp^*c} \leq 1, c \in L^\infty(\R^2)\Big\}.
		\label{eq: strong duality}
	\end{align}
	Moreover, if $\dualGaborProb$ has a solution $c_0 \in L^\infty(\R^2)$, then
	\begin{equation}
		\bigcup_{\nu_0 \in \mathrm{Sol}\{\gaborProb\}} \supp(\nu_0) \subseteq \{t \in \R \colon \abs{(\gaborOp^*c_0)(t)} = 1\}.
		\label{eq: implication dual problem}
	\end{equation}
\end{thm}

Theorem~\ref{thm: fenchel dual gabor prob} follows by application of~\cite[Thms.~4.4.2 and 4.4.3]{Borwein2005} similarly to what was done in~\cite[Prop.~2]{Bredies2012}.
We emphasize that $\dualGaborProb$ is the \textit{pre}dual problem of $\gaborProb$, meaning that $\gaborProb$ is the dual problem of~$\dualGaborProb$. The dual problem of $\gaborProb$ is, however, not $\dualGaborProb$ as the space $L^1(\R^2)$ is not reflexive.

The consequences of Theorem~\ref{thm: fenchel dual gabor prob} are the following: Assuming that $\dualGaborProb$ has a solution, which we denote by $c_0 \in L^\infty(\R^2)$, the support $\measSupport = \{t_\measSpikeIdx\}_{\measSpikeIdx \in \measSupportIdx}$ of the measure $\meas$ to be recovered must satisfy $\abs{(\gaborOp^*c_0)(t_\measSpikeIdx)} = 1$ if $\meas$ is to be in the set of solutions of $\gaborProb$. Furthermore, if $\dualGaborProb$ has a solution $c_0 \in L^\infty(\R^2)$ such that $\abs{\gaborOp^*c_0}$ is not identically $1$ on $\R$, then every solution to $\gaborProb$ is a discrete measure despite the fact that $\gaborProb$ is an optimization problem over the space of \textit{all} complex Radon measures. This can be seen as follows: Both the window function $\window$ and $\gaborOp^*c_0$ can be extended to entire functions which we also denote by $\window$ and $\gaborOp^*c_0$, i.e.,
\begin{align}
	&\forall z \in \C, \quad \window(z) = \frac{1}{\sqrt{\sigma}} \exp\left(-\frac{\pi z^2}{2\sigma^2}\right) \quad \text{and} \notag\\
	&(\gaborOp^*c_0)(z) = \int_{-\freq_c}^{\freq_c}\int_\R c_0(\tau, \freq)\window(z - \tau)e^{2\pi i\freq z}\mathrm{d}\tau\mathrm{d}\freq. \label{eq: interpolating function}
\end{align}
We can then define the function
\begin{equation*}
	\forall z \in \C, \quad h(z) \triangleq 1 - (\gaborOp^*c_0)(z)\overline{(\gaborOp^*c_0)(\overline{z})}.
\end{equation*}
Since $\abs{\gaborOp^*c_0}$ is not identically $1$ on $\R$, $h$ is not identically zero. Consequently, on the basis of~\cite[Thm.~10.18]{Rudin1987} the set $\{z \in \C \colon h(z) = 0\}$, and a fortiori the set $\{t \in \R \colon \abs{\gaborOp^*c_0} = 1\}$, are at most countable and have no limit points. But since \eqref{eq: implication dual problem} holds, this implies that any solution $\nu_0$ to $\gaborProb$ must have discrete support, and therefore, $\nu_0$ is necessarily a discrete measure.

Similarly to~\cite[Sec.~2.4]{Duval2013}, the following theorem provides a necessary and sufficient condition for $\meas$ to be a solution of $\gaborProb$.

\begin{thm}[Optimality conditions]
	\label{thm: equivalence condition mu sol}
	The measure $\meas$ to be recovered is in the set of solutions of $\gaborProb$ if and only if there exists $c_0 \in L^\infty(\R^2)$ such that
	\begin{equation*}
		\normInfty{\gaborOp^*c_0} \leq 1 \quad \text{and} \quad \forall \measSpikeIdx \in \measSupportIdx, \quad (\gaborOp^*c_0)(t_\measSpikeIdx) = \frac{\measAmplitude{\measSpikeIdx}}{\abs{\measAmplitude{\measSpikeIdx}}}.
	\end{equation*}
\end{thm}

It is important to note that $\dualGaborProb$ has at least one solution if the measure $\meas$ to be recovered is a solution of $\gaborProb$, as the $\sup$ in~\eqref{eq: strong duality} is attained for~$c_0$. Indeed, the following holds:
\begin{align*}
	\innerProd{c_0}{\measurements} &= \innerProd{c_0}{\gaborOp\meas} = \innerProd{\gaborOp^*c_0}{\meas} = \int_\group \overline{(\gaborOp^*c_0)(t)} \dmeas{\meas}{t} \\
		&= \sum_{\measSpikeIdx \in \measSupportIdx} \measAmplitude{\measSpikeIdx} \overline{(\gaborOp^*c_0)(t_\measSpikeIdx)} = \sum_{\measSpikeIdx \in \measSupportIdx} \abs{\measAmplitude{\measSpikeIdx}} = \normTV{\meas}.
\end{align*}
Since $\normInfty{\gaborOp^*c_0} \leq 1$, $c_0$ is a solution of $\dualGaborProb$.

Theorem~\ref{thm: equivalence condition mu sol} provides conditions on $\meas$ to be a solution of~$\gaborProb$. However, we hope for more, namely, we want conditions on $\meas$ to be the \emph{unique} solution of $\gaborProb$. Such conditions are given in the following theorem, which is a straightforward adaptation of \cite[App.~A]{Candes2012}.

\begin{thm}[Uniqueness]
	If for every sequence $\varepsilon = \{\varepsilon_\measSpikeIdx\}_{\measSpikeIdx \in \measSupportIdx}$ of unit magnitude complex numbers, there exists a function $c_0 \in L^\infty(\R^2)$ obeying
	\begin{align}
		\forall \measSpikeIdx \in \measSupportIdx, \quad &(\gaborOp^*c_0)(t_\measSpikeIdx) = \varepsilon_\measSpikeIdx \label{eq: interpolation}\\
		\forall t \in \R\!\setminus\!\measSupport, \quad &\abs{(\gaborOp^*c_0)(t)} < 1, \label{eq: constraint interpolation}
	\end{align}
	then $\meas$ is the unique solution of $\gaborProb$.
	\label{thm: uniqueness dual certificate}
\end{thm}

Verifying the conditions of Theorem~\ref{thm: uniqueness dual certificate} requires solving constrained interpolation problems associated with the support set $\measSupport = \{t_\measSpikeIdx\}_{\measSpikeIdx \in \measSupportIdx}$ of $\meas$, as specified by \eqref{eq: interpolation} and \eqref{eq: constraint interpolation}. The following theorem provides conditions ensuring that these solutions can be given in explicit form.

\begin{thm}[Exact recovery]
\label{thm: exact recovery}
	Let $\sigma = \frac{1}{4\freq_c}$. If the minimum distance $\Delta$ between any two points of $\measSupport$,
	\begin{equation*}
		\Delta = \inf_{\substack{\measSpikeIdx, \measSpikeIdx' \in \measSupportIdx \\ \measSpikeIdx \neq \measSpikeIdx'}} \abs{t_\measSpikeIdx - t_{\measSpikeIdx'}},
	\end{equation*}
	satisfies $\Delta > 1/\freq_c\,$,
	then the conditions of Theorem~\ref{thm: uniqueness dual certificate} are met, and, hence, $\meas$ is the unique solution of~$\gaborProb$. 
\end{thm} 

We next briefly describe the main ingredients of the proof of Theorem~\ref{thm: exact recovery}, which is mostly inspired by~\cite[Sec.~2, pp.~15--27]{Candes2012}. The proof is accomplished by verifying the conditions of Theorem~\ref{thm: uniqueness dual certificate}. To this end, we fix a sequence $\varepsilon = \{\varepsilon_\measSpikeIdx\}_{\measSpikeIdx \in \measSupportIdx}$ of complex unit-magnitude numbers, and take a function $c_0 \in L^\infty(\R^2)$ of the form
\begin{equation*}
	c_0(\tau, \freq) \!\triangleq \!\sum_{\measSpikeIdx \in \measSupportIdx} \!\left[\alpha_\measSpikeIdx \window(t_\measSpikeIdx - \tau)e^{-2\pi i\freq t_\measSpikeIdx} \!\!+\! \beta_\measSpikeIdx \window'(t_\measSpikeIdx - \tau)e^{-2\pi i\freq t_\measSpikeIdx}\right]
\end{equation*}
for $(\tau, \freq) \in \R^2$, where $\alpha = \{\alpha_\measSpikeIdx\}_{\measSpikeIdx \in \measSupportIdx}$ and $\beta = \{\beta_\measSpikeIdx\}_{\measSpikeIdx \in \measSupportIdx}$ are sequences in $\ell^\infty(\measSupportIdx)$. The function $\gaborOp^*c_0$ is thus given by
\begin{align}
	\forall t \in \R, &\quad (\gaborOp^*c_0)(t) = \sum_{\measSpikeIdx \in \measSupportIdx} \alpha_\measSpikeIdx \autocorrelation(t - t_\measSpikeIdx)\sinc(2\pi\freq_c(t - t_\measSpikeIdx)) \notag\\
	&+ \sum_{\measSpikeIdx \in \measSupportIdx} \beta_\measSpikeIdx \autocorrelation'(t_\measSpikeIdx - t)\sinc(2\pi\freq_c(t - t_\measSpikeIdx)), \label{eq: dual certificate gabor prob}
\end{align}
where $\autocorrelation$ designates the autocorrelation of $\window$, that is,
\begin{equation*}
	\forall t \in \R, \quad \autocorrelation(t) = \int_{\R} \window(\tau)\window(t + \tau)\mathrm{d}\tau = \exp\left(-\frac{\pi t^2}{4\sigma^2}\right).
\end{equation*}
Next, we determine sequences $\alpha$ and $\beta$ such that the interpolation conditions $(\gaborOp^*c_0)(t_\measSpikeIdx) = \varepsilon_\measSpikeIdx$ are satisfied and $\gaborOp^*c_0$ has a local extremum at every $t_\measSpikeIdx$, $\measSpikeIdx \in \measSupportIdx$.
The coefficients $\alpha_\measSpikeIdx$ play the main role in interpolating the points $(t_\measSpikeIdx, \varepsilon_\measSpikeIdx)$. The coefficients $\beta_\measSpikeIdx$ add a correction term to ensure that $\gaborOp^*c_0$ does not exceed $1$ on $\R$. To conclude the proof, we finally show that \eqref{eq: constraint interpolation} holds.

We can see from~\eqref{eq: dual certificate gabor prob} how the windowing in the STFT helps to improve the recovery guarantee in Theorem~\ref{thm: exact recovery} compared to recovery based on unwindowed Fourier measurements, as considered in~\cite[Sec.~2]{Candes2012}. Specifically, in the case of unwindowed Fourier measurements, the interpolation function $\mathcal{A}^*c_0$ must be a Paley-Wiener function~\cite[Thm.~19.3]{Rudin1987}, while here $\gaborOp^*c_0$ is clearly not band-limited due to Gaussian windowing and therefore has better time-localization. This, in turn, allows the minimum separation $\Delta$ to be smaller.

\section{Simulations}
\label{section: simulations}

For the simulation results we consider the recovery of the discrete complex measure $\meas = \sum_{\measSpikeIdx \in \measSupportIdx} \measAmplitude{\measSpikeIdx} \delta_{t_\measSpikeIdx} \in \measSpace{\torus}$ over the torus $\torus$ (the set $\measSupportIdx$ is then finite). The Gaussian window function is periodized so that for all $t \in \R$,
\begin{equation}
	\window(t) = \sum_{n \in \Z} \frac{1}{\sqrt{\sigma}} \exp\left(-\frac{\pi(t + n)^2}{2\sigma^2}\right) = \sum_{n \in \Z} \window_n e^{2\pi int},
	\label{eq: window torus}
\end{equation}
with $\window_n \triangleq \sqrt{2\pi} \exp\left(-2\pi\sigma^2n^2\right)$,  $n \in \Z$.
The dual group of~$\torus$ is $\Z$. The corresponding STFT measurements of $\meas$ are given by the sequence $\{\measurements_k\}_{k = -\freq_c}^{\freq_c}$ of functions
\begin{equation*}
	\forall \tau \in \torus, \!\!\!\quad \measurements_k(\tau) \triangleq \!\sum_{\measSpikeIdx \in \measSupportIdx} \!\measAmplitude{\measSpikeIdx}\window(t_\measSpikeIdx - \tau) e^{-2\pi ikt_\measSpikeIdx} \!=\!\sum_{n \in \Z} \measurements_{k, n}e^{2\pi in\tau}\!,
\end{equation*}
where $\measurements_{k, n} \triangleq \sum_{\measSpikeIdx \in \measSupportIdx} \measAmplitude{\measSpikeIdx} \window_n e^{-2\pi i(n+k)t_\measSpikeIdx}$ is the $n$th Fourier coefficient of $\measurements_k$. Using Parseval's theorem, the objective function for $\dualGaborProb$ can be rewritten as
\begin{equation*}
	\innerProd{c}{\measurements} = \sum_{k = -\freq_c}^{\freq_c}\sum_{n \in \Z} c_{k,n}\measurements_{k,n},
\end{equation*}
where $c_{k,n}$ denotes the $n$th Fourier coefficient of $c_k \in L^\infty(\torus)$ for $\abs{k} \leq \freq_c$.
The function $\gaborOp^*c$ can be expressed as
\begin{equation*}
	\forall t \in \torus, \quad (\gaborOp^*c)(t) = \sum_{k = -\freq_c}^{\freq_c}\sum_{n \in \Z} \window_nc_{k, n}e^{2\pi i(k+n)t}.
\end{equation*}
In order to render the problem $\dualGaborProb$ finite-dimensional, we approximate the function $\window$ in~\eqref{eq: window torus} by keeping only $2N+1$ Fourier coefficients, that is, we replace $\window$ by 
\begin{equation*}
	\forall t \in \torus, \quad \widetilde{\window}(t) =  \sum_{n = -N}^{N} \window_n e^{2\pi int},
\end{equation*}
where $N$ is chosen large enough for the coefficients $\window_n$, $\abs{n} \geq N$, to be small. The  objective of $\gaborProb$ then is $\innerProd{\mathbf{C}}{\mathbf{Y}}$, where $\mathbf{C} \triangleq (c_{k, n})_{\abs{k} \leq \freq_c, \abs{n} \leq N}$ and $\mathbf{Y} \triangleq (\measurements_{k, n})_{\abs{k} \leq \freq_c,\abs{n} \leq N}$.
The function $\gaborOp^*c$ becomes a trigonometric polynomial which can be expressed as
\begin{equation*}
	\sum_{m = -(\freq_c+N)}^{\freq_c+N} x_me^{2\pi imt} \quad \text{with} \quad x_m = \sum_{n = n_\mathrm{min}}^{n_\mathrm{max}} \window_mc_{m -n, n},
\end{equation*}
where $n_\mathrm{min} \triangleq \max\{-N, m-\freq_c\}$ and $n_\mathrm{max} \triangleq \min\{N, \freq_c+m\}$. We can now apply a procedure similar to the one developed in \cite[Sec.~4, pp.~31--36]{Candes2012} to solve $\dualGaborProb$ and to reconstruct the corresponding solution of $\gaborProb$.

\definecolor{fourierColor}{RGB}{153, 8, 247}
\definecolor{stft10Color}{RGB}{8, 208, 247}
\definecolor{stft5Color}{RGB}{9, 140, 247}
\definecolor{stft05Color}{RGB}{0, 0, 250}

\begin{figure}
	\includegraphics[width = \columnwidth]{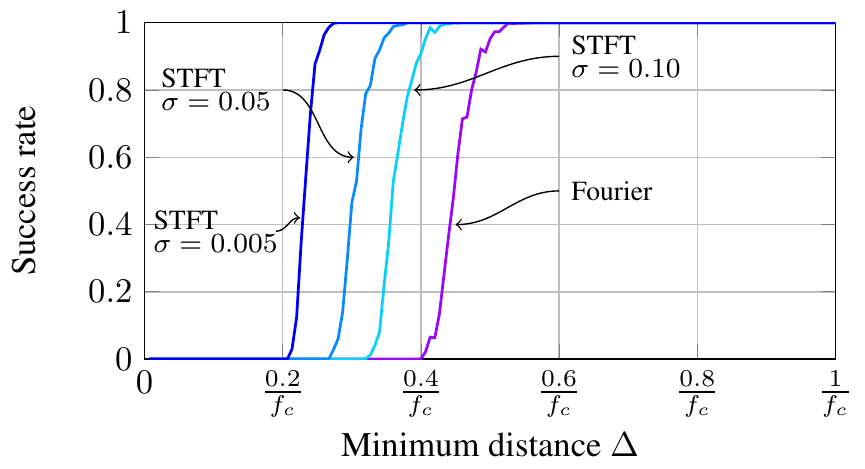}
	\caption{Success rate for support recovery from (unwindowed) Fourier measurements and from STFT measurements with $\freq_c = 50$ and $N = 50$.}
	\label{fig: monte carlo simulations for the support}
\end{figure}

To assess the performance of our recovery procedure, we run $1500$ trials. For each $\Delta$, we construct a discrete Radon measure $\meas$ supported on the set $\measSupport = \{t_\measSpikeIdx\}_{\measSpikeIdx = 0}^{S}$ with $S = \lfloor 1/(2\Delta)\rfloor$ and $t_\measSpikeIdx = 2\measSpikeIdx\Delta + r_\measSpikeIdx$, where $r_\measSpikeIdx$ is chosen uniformly at random in $[0, \Delta]$. Thus, the minimum distance between two distinct points of $\measSupport$ is at least $\Delta$. The complex amplitudes are obtained by choosing their real and imaginary parts uniformly at random in $[0, 1000]$. If the reconstructed measure $\hat{\meas}$ has support $\hat{T} = \{\hat{t}_\measSpikeIdx\}_{\measSpikeIdx \in \measSupportIdx}$ satisfying $\|\hat{T} - T\|_{\ell^2}/\|T\|_{\ell^2} \leq 10^{-3}$, we declare success. The corresponding results are depicted in Fig.~\ref{fig: monte carlo simulations for the support}. As predicted by our theoretical results, we, indeed, observe a factor-of-two improvement in the case of recovery from STFT measurements relative to recovery from unwindowed Fourier measurements as in~\cite[Sec.~2]{Candes2012}. Note, however, that time-frequency measurements provide more information than frequency-only measurements as considered in~\cite{Candes2012}.

\bibliographystyle{IEEEtran} 
\bibliography{ref}

\end{document}